\documentclass[10pt]{IEEEtran}

\usepackage{amsmath,amssymb}
\usepackage[boxed,linesnumbered]{algorithm2e}
\usepackage{graphicx,epsfig,url}
\usepackage{microtype}
\usepackage[caption=false,font=footnotesize]{subfig}
\usepackage[usenames, dvipsnames]{color}

\def\b{\mathbf{b}}

\def\A{\mathbf{A}}

\def\i{\boldsymbol{i}}
\def\j{\boldsymbol{j}}

\def\c{\boldsymbol{c}}

\ifCLASSINFOpdf
\else
\fi

\newtheorem{theorem}{Theorem}
\newtheorem{proposition}[theorem]{Proposition}

\begin{document}

\markboth{IEEE SIGNAL PROCESSING LETTERS, VOL. 25, NO. 10, OCTOBER 2018}{}

\title{Optimized Fourier Bilateral Filtering}

\author{Sanjay Ghosh,~\IEEEmembership{Student~Member,~IEEE}, Pravin Nair,~\IEEEmembership{Student~Member,~IEEE}, and Kunal N. Chaudhury,~\IEEEmembership{Senior~Member,~IEEE} \thanks{Address: Department of Electrical Engineering, Indian Institute of Science, Bangalore 560012, India. Correspondence: sanjayg@iisc.ac.in.}}

\maketitle

\begin{abstract}
We consider the problem of approximating a truncated Gaussian kernel using Fourier (trigonometric) functions.
The computation-intensive bilateral filter can be expressed using fast convolutions by 
applying such an approximation to its range kernel, where the truncation in question is the dynamic range of the input image. 
The error from such an approximation depends on the period, the number of sinusoids, and the coefficient of each sinusoid. 
For a fixed period, we recently proposed a model for optimizing the coefficients using least-squares fitting. Following the Compressive Bilateral Filter (CBF), we demonstrate that the approximation can be improved by taking the period into account during the optimization.  {The accuracy of the resulting filtering is found to be at least as good as CBF, but significantly better for certain cases. The proposed approximation can also be used for non-Gaussian kernels, and it comes with guarantees on the filtering accuracy.}
 \end{abstract}


\section{Introduction}

The bilateral filter is popularly used in image processing and computer vision for edge-preserving smoothing \cite{Tomasi1998,Paris2009}. Unlike classical convolutional filters, it uses an additional kernel for measuring proximity in range (intensity) space. In particular, the bilateral filtering of an image $f$ is given by
\begin{equation}
\label{BF}
 f_{\mathrm{BF}}(\i)=  \frac{\sum_{\j} \omega(\j) \  \varphi\big(f(\i-\j)-f(\i) \big)  f(\i-\j)}{\sum_{\j} \omega(\j)    \varphi\big(f(\i-\j)-f(\i) \big)  },
\end{equation}
where $\omega$ and $\varphi$ are the spatial and range kernels. Both kernels are typically Gaussian \cite{Tomasi1998}:
\begin{equation}
\label{denom}
\omega(\i) = \exp\left(- \frac{\lVert \i \rVert^2}{2\theta^2}\right) \quad \text{and} \quad \varphi(t) = \exp\left(- \frac{t^2}{2\sigma^2}\right),
\end{equation}
where $\theta$ and $\sigma$ are the respective standard deviations.

Notice that the range kernel acts on the intensity difference between the pixel of interest $\i$ and its neighbor $\i-\j$. If $|f(\i)-f(\i-\j)| \gg3\sigma$ (e.g. the pixels are on opposite sides of an edge), then the weight assigned to pixel $\i-\j$ is small, whereby it is excluded from the aggregation. This helps in preserving sharp edges \cite{Tomasi1998}. However, the range kernel also makes the filter non-linear and computation intensive. In particular, the brute-force computation of \eqref{BF} requires $2(6\theta+1)^2$ operations per pixel. This makes the real-time implementation challenging for practical values of $\theta$. Several fast approximations \cite{Durand2002}--\cite{Nair2017} have been proposed to accelerate the brute-force computation of \eqref{BF}. 
The present focus is on approximations that use trigonometric functions \cite{Chaudhury2011a,Kamata2015,GhoshSPL2016,Nair2017}. We briefly describe their relation to the present work and summarize our contributions.

\begin{figure*}
\centering
\subfloat[Approximation error vs. $T$.]{\includegraphics[width=0.23 \linewidth]{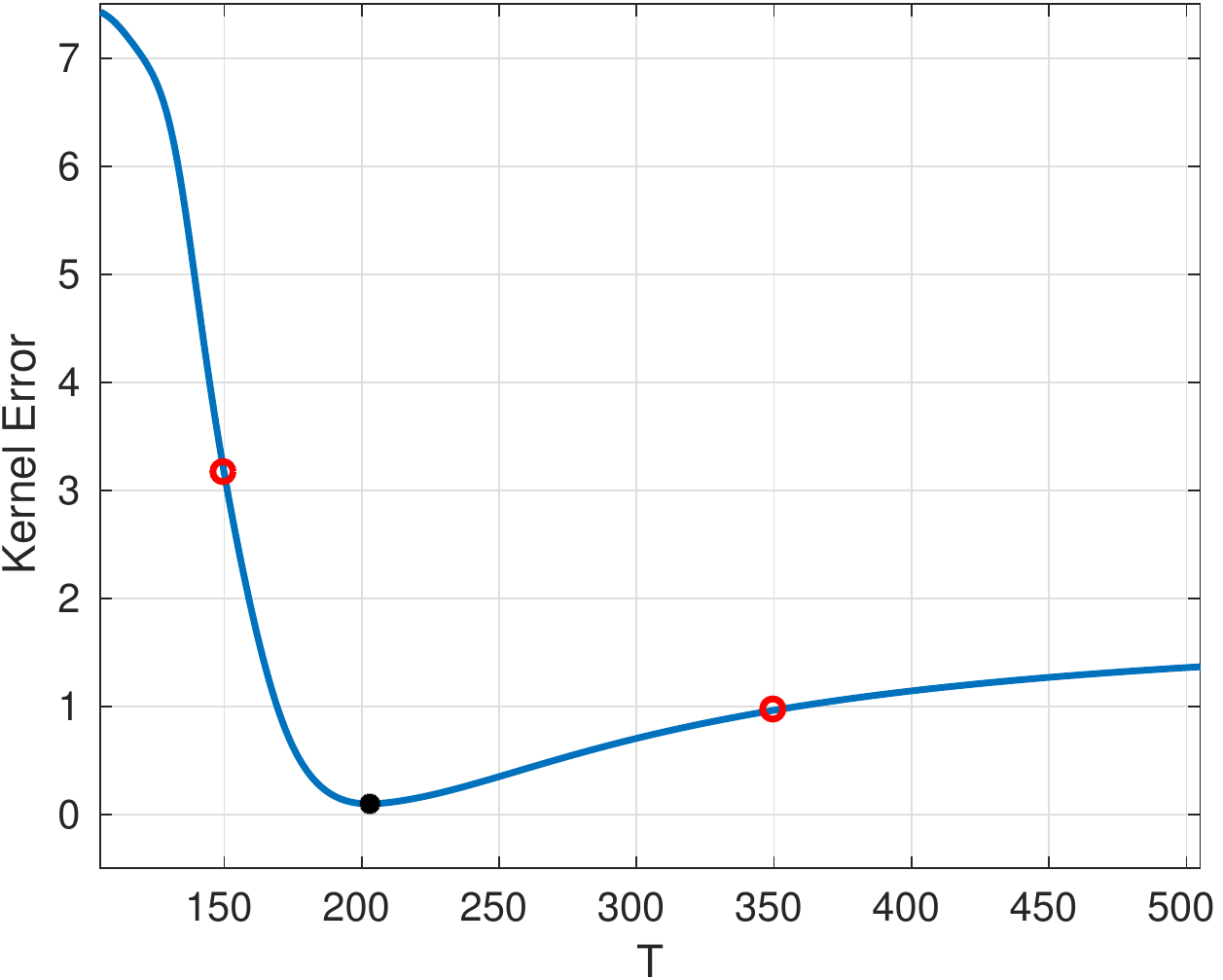}} \hspace{-0.5mm}
\subfloat[$T = 203$.]{\includegraphics[width= 0.24 \linewidth]{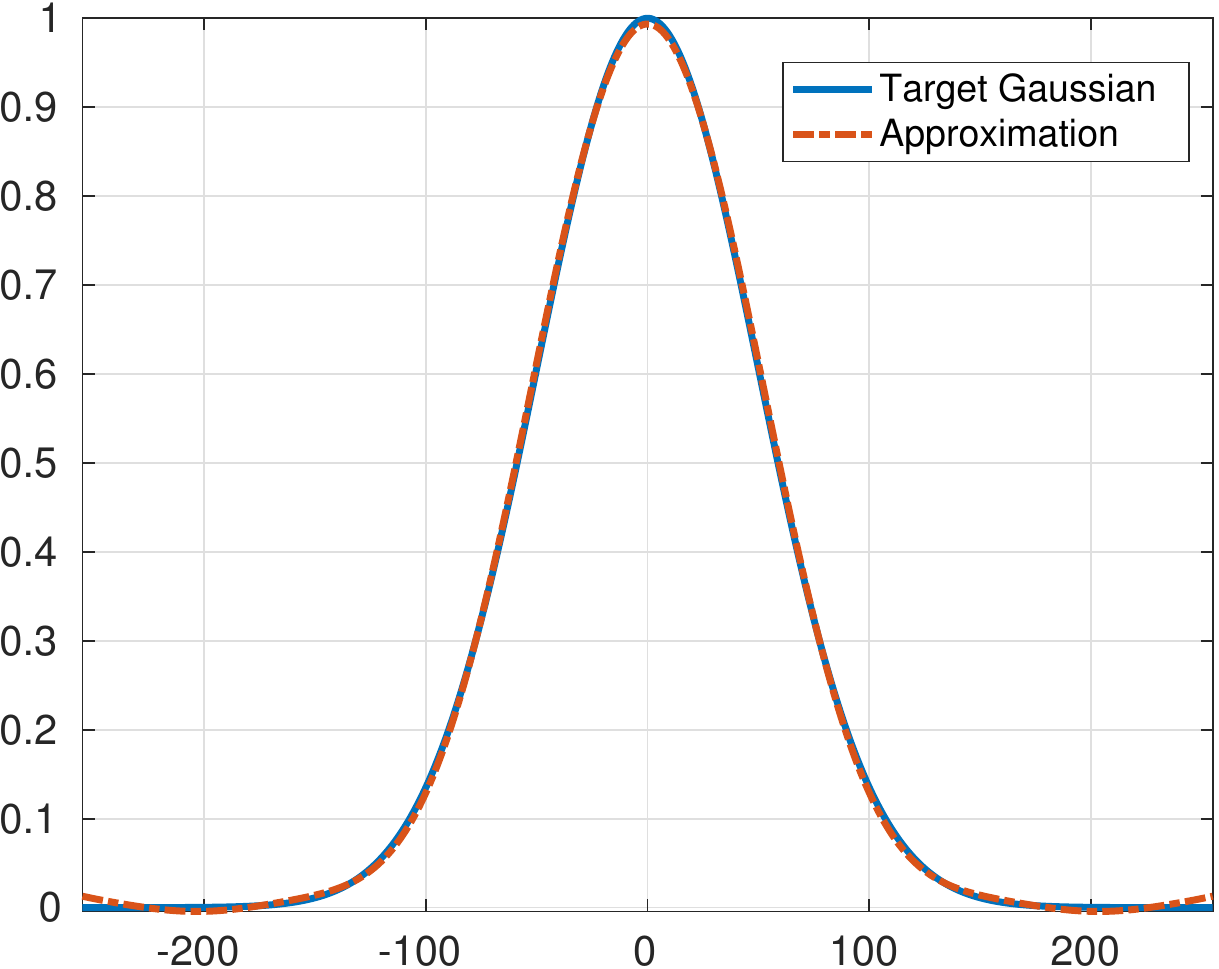}} \hspace{-0.5mm}
\subfloat[$T = 150$.]{\includegraphics[width=0.24 \linewidth]{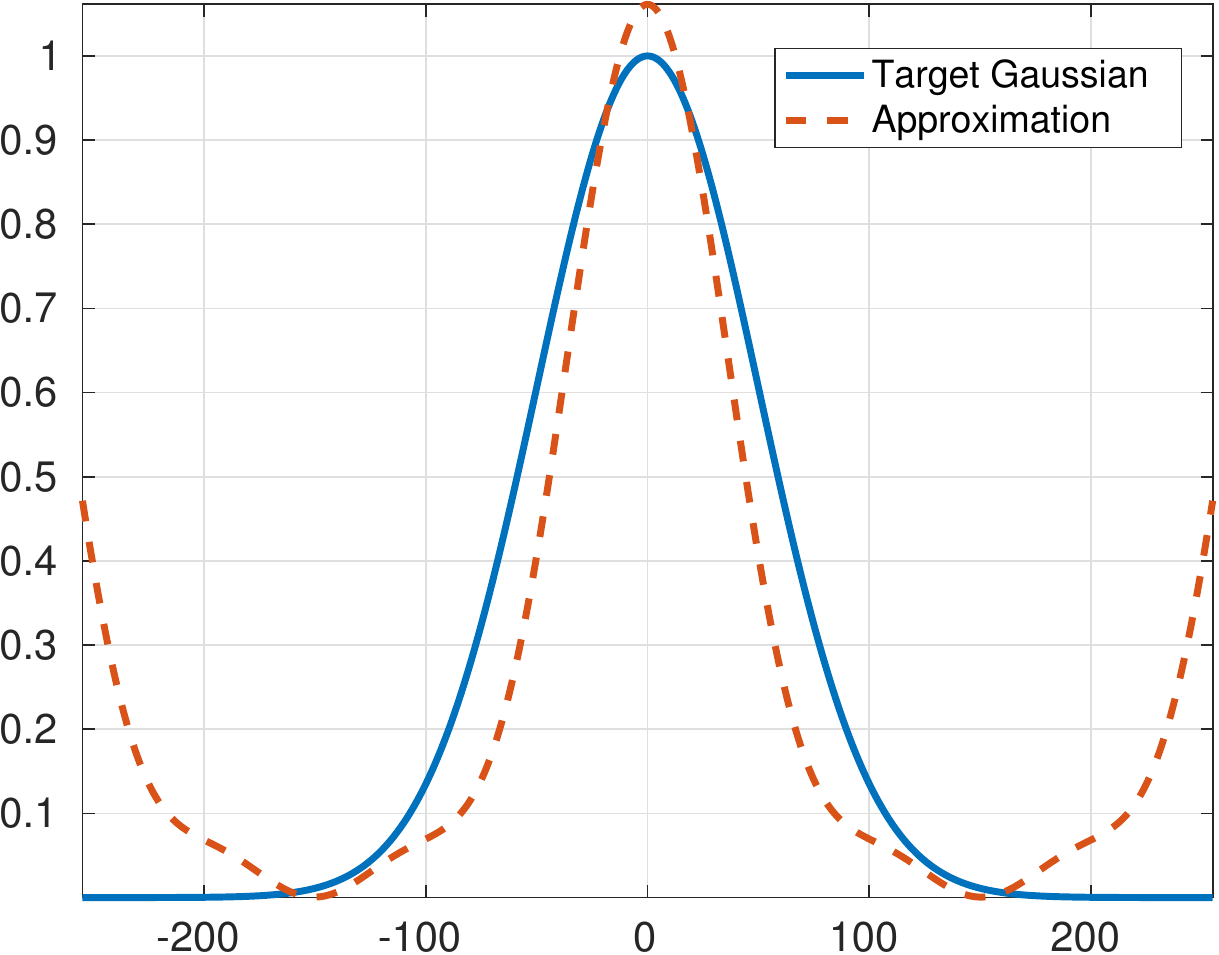}} \hspace{-0.5mm}
\subfloat[$T = 350$.]{\includegraphics[width= 0.24 \linewidth]{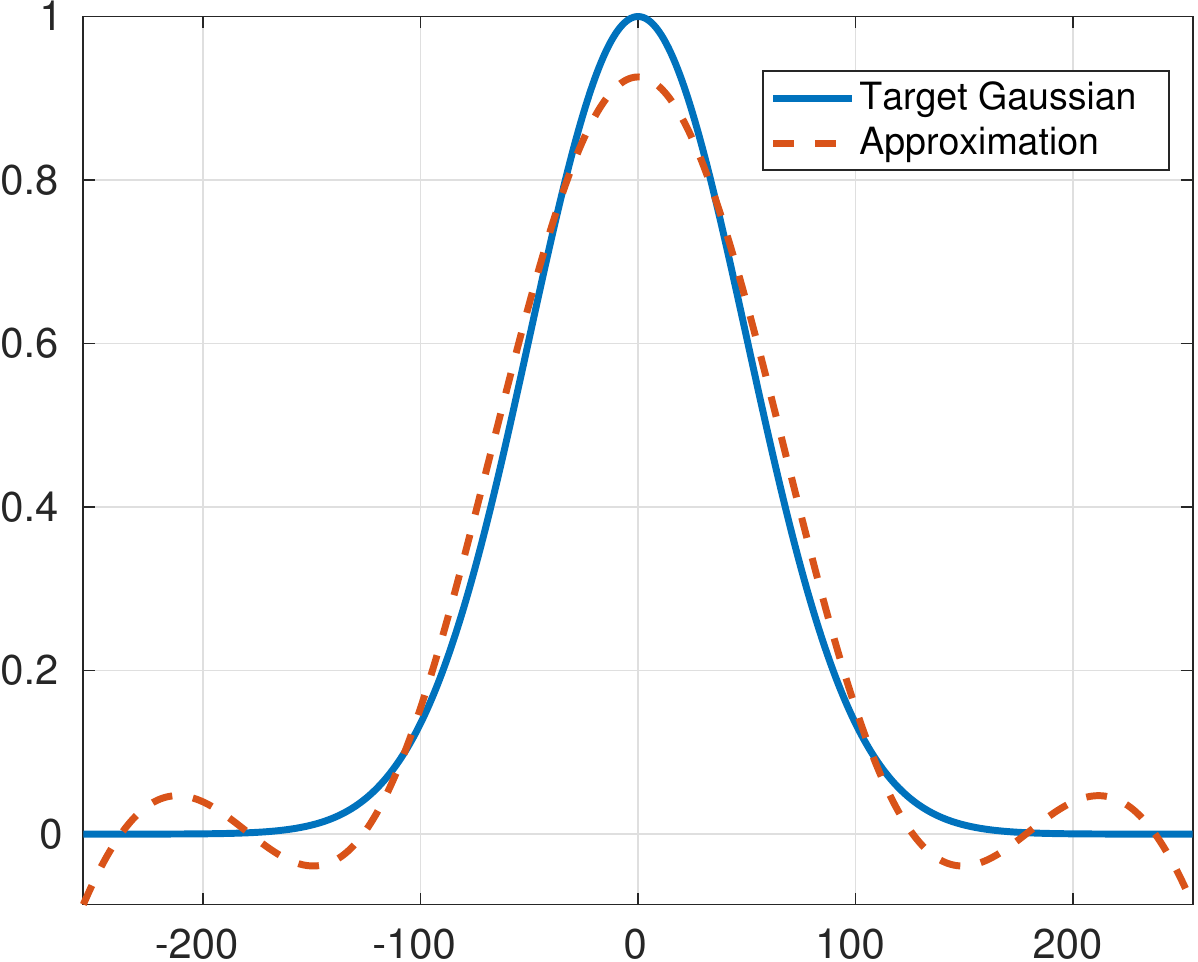}}
\caption{Variation of the kernel error $E(K,T)$ with $T$ for fixed $K = 4$. For this example, $R=255, \sigma = 50,$ and $\varepsilon = 0.1$. Notice in (a) that the error is minimum when $T = 203$. The approximation for this value of $T$ is shown in (b), along with the target Gaussian. The kernel approximations for non-optimum values of $T$ are shown in (c) and (d), which are visibly poor (the corresponding errors are marked with red circles in (a)).}
\label{fig:Err_vs_T}
\end{figure*} 

\textbf{Previous work}. The idea of fast bilateral filtering using trigonometric functions was first proposed in \cite{Chaudhury2011a}.
It was observed that, by approximating $\varphi$ using trigonometric polynomials, we can compute \eqref{BF} using fast convolutions.
This was later refined and extended in \cite{Kamata2015}--\cite{Nair2017}. In particular, state-of-the-art results were obtained using Fourier series approximation of a truncated $\varphi$ in \cite{Kamata2015}, where the truncation is simply the intensity range of the input image. An important observation from \cite{Kamata2015}, which was overlooked in prior work \cite{GhoshSPL2016}, is that the period of the sinusoids plays a critical role in the approximation. In \cite{GhoshSPL2016}, the period was set as $[-R,R]$, where $R$ is the intensity range. While this works well when $\sigma$ is small, it results in poor approximations when $3\sigma \gg R$. The kernel does not flatten out sufficiently over $[-R,R]$ in such cases. This induces (higher-order) discontinuities at the boundaries, causing the coefficients to decay slowly (e.g. see Figure $4$ in \cite{Kamata2015}). 

\textbf{Contributions}. In this work, we combine some of the ideas from \cite{Kamata2015} and \cite{GhoshSPL2016} to develop an approximation with improved filtering quality. By filtering quality, we mean the error between \eqref{BF} and the output obtained using the kernel approximation in question. The highlights of our approximation model and its key differences with \cite{Kamata2015} are as follows:\\
\indent $\bullet$  {Unlike \cite{Kamata2015}, we do not use a continuous approximation of $\varphi$ over the interval $[-R,R]$. Instead, we just approximate the discrete samples $\varphi(t)$ corresponding to the intensity levels $t \in \{-R,\ldots,R\}$. Notice that it is precisely these samples that appear in \eqref{BF} and \eqref{denom}. As a result, we can control the filtering quality (see Proposition \ref{prop1}) by adjusting the kernel error at will.}\\
\indent  $\bullet$  {In \cite{Kamata2015}, the Gaussian kernel is approximated using Fourier series. However, since there is no known analytical formula for the Fourier coefficients in this case, the coefficients are further approximated using the Fourier transform of a Gaussian (cf. \cite[Lemma 8]{Kamata2015}). In contrast, the coefficients in our model are computed exactly using least-squares optimization}.\\
\indent $\bullet$  {The approximation error for our method provably decreases with the increase in the number of sinusoids ($K$), and eventually vanishes (see Proposition \ref{prop2}). Such a simple guarantee is not offered by \cite{Kamata2015}. This is because the period changes with $K$ in \cite{Kamata2015}. 
Therefore, we cannot use standard convergence guarantees from Fourier analysis (where the period is assumed to be fixed)}.\\
\indent $\bullet$  {The approximation in \cite{Kamata2015} is specialized for Gaussians, whereas our approximation can be used for any  range kernel. As is well-known, it is rather difficult to analytically compute the Fourier coefficients (or the Fourier transform) of an arbitrary kernel. As a result, it is difficult to use the integration-based method  in \cite{Kamata2015} for computing the coefficients for non-Gaussian kernels \cite{Durand2002,Mirbach2012, Yang2014}. In other words, the issue with \cite{Kamata2015} for non-Gaussian kernels is the fast and accurate computation of Fourier coefficients.}

\textbf{Organization}. The rest of the paper is organized as follows. The proposed approximation, along with some numerical results, is explained in Section \ref{PO}. In Section \ref{RC}, we compare our filtering results with \cite{Kamata2015,GhoshSPL2016}, both visually and in terms of PSNR. We summarize the results in Section \ref{Conc}.

\section{Parameter Optimization}
\label{PO}

Following \cite{Kamata2015}, consider the $K$-term Fourier series approximation of $\varphi$ over the period $[-T,T]$:
\begin{equation}
\label{approx}
\hat{\varphi}(t) = \sum_{k=0}^{K-1} c_k \cos\left(\frac{2\pi k t}{2T+1}  \right).
\end{equation}
Since the original kernel $\varphi$ is symmetric, we consider only the cosine terms in \eqref{approx}. Notice that, unlike the standard definition, we divide by $2T+1$ and not by $2T$ in \eqref{approx}. The reason for this deviation will be made precise later. The important observation is that, by using $\hat{\varphi}$ in place of $\varphi$, we can express the numerator and denominator of \eqref{BF} using spatial convolutions; see Section II in \cite{GhoshSPL2016} for a detailed account. This observation is at the heart of the fast algorithms in \cite{Kamata2015,GhoshSPL2016}.

For fixed $\sigma$, the design problem is that of fixing the intrinsic parameters $K, T$ and $(c_k)$. More precisely, since the number of convolutions required in the fast algorithm is proportional to $K$, the goal is to find the smallest $K$ (and the corresponding $T$ and $(c_k)$) such that the error between $\varphi$ and $\hat{\varphi}$ is within a specified tolerance \cite{Kamata2015,GhoshSPL2016}.

A key difference between the proposals in \cite{Kamata2015,GhoshSPL2016} is the definition of approximation error. 
Notice that the argument $t$ in $\varphi(t)$ assumes the values $f(\i) - f(\i-\j)$ in \eqref{BF}. 
Thus, if the intensity of the input image is in the range $[0,R]$, then $t$ takes values in $[-R,R]$.
The error (up to a normalization) was defined in \cite{Kamata2015} to be
\begin{equation*}
\int_{-R}^R \big( \varphi(t) - \hat{\varphi}(t) \big)^2 \ dt.
\end{equation*}
However, notice that the domain of $t$ is not the full interval $[-R,R]$, but rather just the integers $\Lambda=\{-R,\ldots,R\}$.
Based on this observation, the following error was considered in \cite{GhoshSPL2016}:
\begin{equation}
\label{error}
 \sum_{t \in \Lambda} \big( \varphi(t) - \hat{\varphi}(t) \big)^2. 
\end{equation}
We choose to work with \eqref{error} for couple of reasons. First, it was shown in \cite{GhoshSPL2016} that a bound on \eqref{error} automatically translates into a bound on the filtering accuracy.
\begin{proposition} 
\label{prop1}
The pixelwise error between the images obtained using $\varphi$ and $\hat{\varphi}$ as range kernels is at most $2R \varepsilon/(\omega(0) -\varepsilon)$, where $0 < \varepsilon < \omega(0)$ is a bound on \eqref{error}. In particular, the pixelwise error vanishes as \eqref{error} becomes small.
\end{proposition}

	The other point is that \eqref{error} is a quadratic function of the coefficients.  {Consider the vector $\b$ of length  $2R+1$ consisting of the samples $\{\varphi(t): t \in \Lambda\}$, and matrix $\A$ of size $(2R+1) \times K$ whose columns are the discretized sinusoids in \eqref{approx}. That is, $\b(i) = \varphi(i - R-1)$ and $\A(i, j) = \cos (\nu (i - R-1)(j-1))$ for $i=1,\ldots, 2R+1$ and $j=1,\ldots,K$, where $\nu=2\pi/(2T+1)$.} We can then simply write \eqref{error} as $\lVert \A \c - \b \rVert^2$, where $\c=(c_k)$. Importantly, for fixed $K$ and $T$, we can exactly minimize \eqref{error} with respect to $\c$ using linear algebra. In particular, let
\begin{equation}
\label{L2error}
E(K,T) = \underset{\c}{\min} \ \lVert \A \c - \b \rVert^2,
\end{equation}
which is the smallest possible error for fixed $K$ and $T$. Notice that the size of $\A$ and its components depend on $K$ and  $T$.

The next question is how does $E(K,T)$ behave with $T$ for some fixed $K$? Intuitively, it is clear that the error is large if $T$ is too small or too large with respect to $\sigma$. If $T$ is too large, then we see from \eqref{approx} that the sinusoids effectively degenerate to constant functions over $[-3\sigma,3\sigma]$.
On the other hand, as mentioned previously, if $T \ll 3\sigma$, then the Gaussian does not flatten out sufficiently within the period $[-T,T]$. This induces a discontinuity at the boundary after the periodization, which causes the Fourier coefficients to decay slowly. This adversely affects the approximation for a fixed $K$. We noticed that this problem persists even after optimizing the coefficients (see Figure \ref{fig:Err_vs_T}).
Thus, following \cite{Kamata2015}, we propose to optimize \eqref{L2error} with $T$. In particular, let
\begin{equation}
\label{eK}
e(K) = \underset{T \in \mathbb{N}}{\min} \ E(K,T).
\end{equation}
Given some user-defined tolerance $\varepsilon >0$, the goal is to find the smallest $K$ such that $e(K) \leq \varepsilon$. The existence of such a $K$ is guaranteed by the following observation (see \textit{supplement} for the proof; this is exactly where we require the division by $2T+1$ in \eqref{approx}).
\begin{proposition}
\label{prop2}
The error given by \eqref{eK} is non-increasing in $K$, i.e., $e(K+1) \leq e(K)$, and it vanishes when $K=2R+1$.  
\end{proposition}

The overall optimization procedure is summarized in Algorithm \ref{algo}, where the optimal order and period are denoted by $K^\star$ and $T^\star$. Proposition \ref{prop2} guarantees that the ``condition $e \leq \varepsilon$'' in line \ref{exit} is satisfied for some $K$. As with the example in Figure \ref{fig:Err_vs_T}, we found that $E(K,T)$ is unimodal in $T$ for fixed $K$. Thus, using a large $T_{\mathrm{max}}$ in line \ref{loopT}, we can obtain the optimal $T$. 
\IncMargin{1mm}
\begin{algorithm}
\KwData{Range parameter $\sigma$ and error tolerance $\varepsilon$.}
\KwResult{Optimal order $K^\star$ and period length $T^\star$.}
Set $K = 0$ and $e = +\infty$\;
 \While{$e > \varepsilon$}{ \label{exit}
 $K = K + 1$\;
 \For{$T=1,2,\ldots,T_{\mathrm{max}}$}{ \vspace{0.5mm}  \label{loopT}
Compute $e_0 = \underset{\c}{\min} \ E(K,T,\c) $\;
\If{$e_0 \leq \min(e,\varepsilon)$} {$K^\star=K, T^\star=T, e=e_0$\;}
 }
 }
\caption{Computation of optimal parameters.}
\label{algo}
\end{algorithm}

Unlike \cite{Kamata2015,GhoshSPL2016}, we propose to perform the optimization offline for practical values of $\sigma$ and $\varepsilon$, and store the corresponding $K^\star$ and $T^\star$  in a lookup table. At run time, we simply need to read $K^\star$ and $T^\star$ from the table, and perform the least-squares optimization in \eqref{L2error} to get the optimal coefficients. 
We empirically found that for fixed $\sigma$, $K^\star$ and $T^\star$ scale almost linearly with $\log(1/\varepsilon)$. Whereas, for fixed $\varepsilon$, $K^\star$ (resp. $T^\star$) decreases (resp. increases)  almost linearly with $\sigma$. Additional plots are provided in the \textit{supplement}. As shown in Figure \ref{fig:OptVal}, $K^\star$ and $T^\star$ varies smoothly with $\sigma$ and $\log(1/\varepsilon)$, and hence the off-grid values can accurately be estimated using bilinear interpolation. The loss in filtering accuracy owing to the sub-optimality of the interpolated $K$ and $T$ is at most $1\mbox{-}2$ dB. For reference, the values of $K^\star$ and $T^\star$ for different values of $\sigma$ and $\varepsilon$ are provided in the file \texttt{LUT.mat} in the supplementary material. Of course, we can use a large table with more entries depending on the application at hand. On the other hand, for hardware implementations of the filter, the values of $\sigma$ and $\varepsilon$ would typically be hardcoded \cite{Gabiger2014}, and a lookup table would not be required.

\begin{figure}
\centering
\subfloat[Optimal $K$.]{\includegraphics[width=0.45 \linewidth]{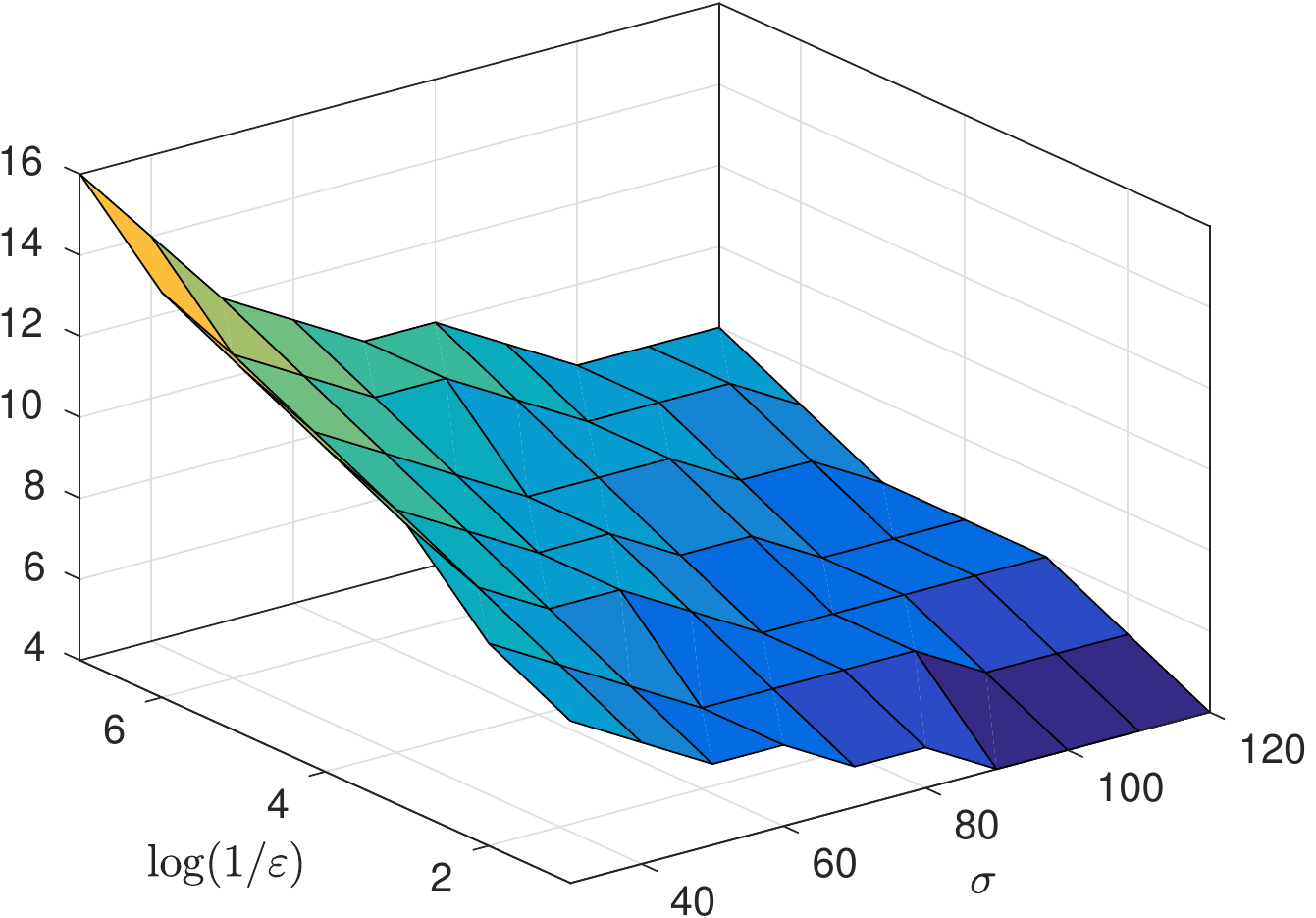}} \hfill
\subfloat[Optimal $T$.]{\includegraphics[width= 0.45 \linewidth]{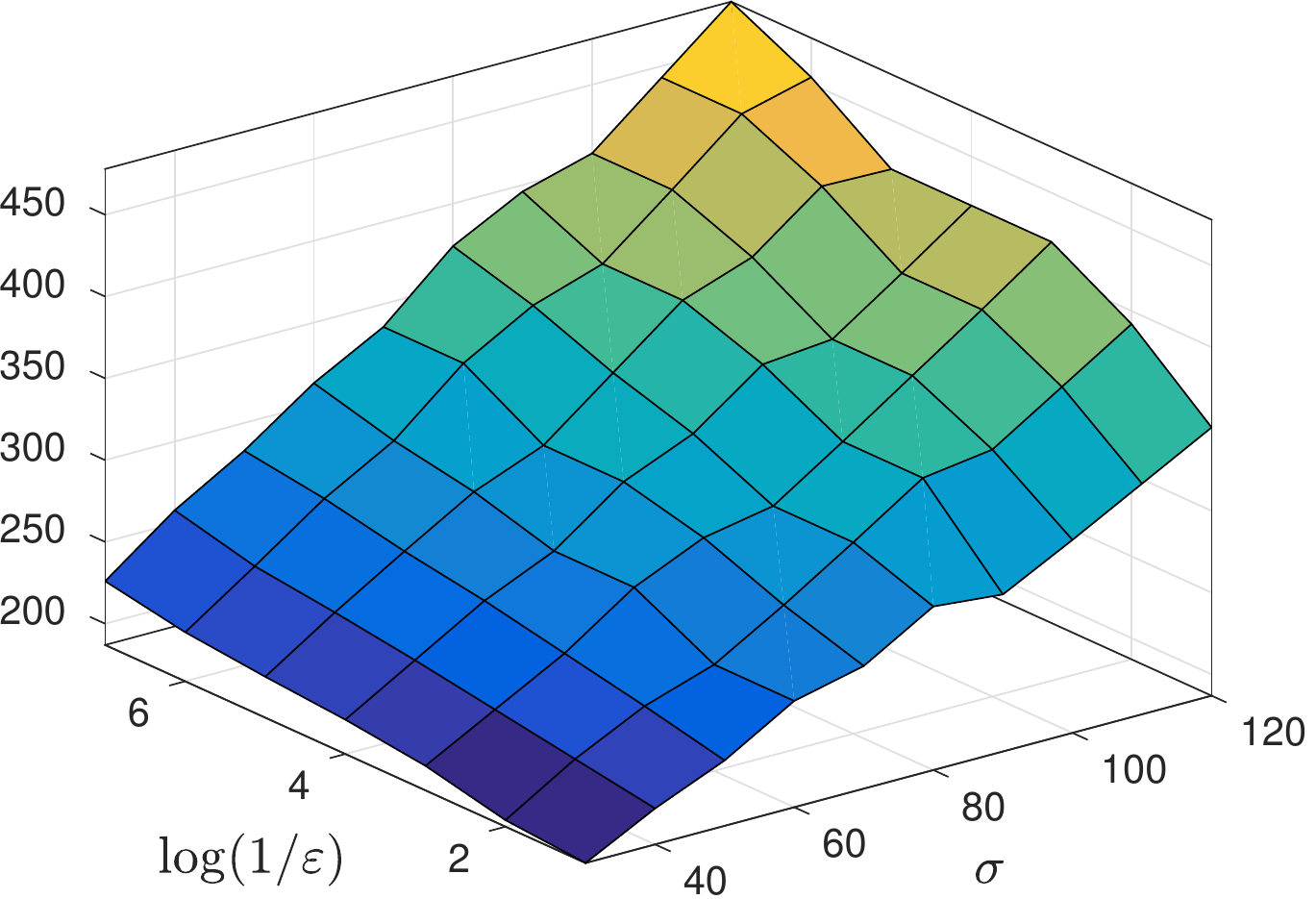}}
\caption{Optimal $K$ and $T$ for different $\sigma$ and $\varepsilon$ obtained using Algorithm \ref{algo}. Notice the smooth trend in either case (see \textit{supplement} for additional plots).}
\label{fig:OptVal}
\end{figure}

\section{Results and Comparisons}
\label{RC}

The simulations in this section were performed using Matlab on a $3.40$ GHz quad-core machine with $32$ GB memory.
We used the Matlab code of \cite{Kamata2015,GhoshSPL2016} provided by the authors.
The Gaussian convolutions involved in the fast algorithm were implemented using the Matlab routine ``imfilter''.
 {We have used $8$-bit grayscale images \cite{ImgDatabase_bm3d}, for which $R=255$.}

\begin{table}
\centering
\caption{Timing and PSNR for the image in Figure \ref{Visualfig1}.}
\label{table}
\begin{tabular}{|c|| c c c ||c c c|}
 \hline
$\sigma$ \textbackslash{} $\varepsilon$   &$1\mathrm{e}\mbox{-}1$  & $1\mathrm{e}\mbox{-}3$ & $1\mathrm{e}\mbox{-}5$  &  $1\mathrm{e}\mbox{-}1$  &  $1\mathrm{e}\mbox{-}3$ &  $1\mathrm{e}\mbox{-}5$  \\ \hline \hline
& \multicolumn{3}{c||}{Timing (ms)}  & \multicolumn{3}{c|}{PSNR (dB)} \\ \hline
$15$   &143 &208  &261     &74.7  &119.4   &166.7 \\ \hline
$30$   &109 &148  &173     &91.4 &140.1   &168.9 \\ \hline
$50$   &79  &109  &128     &91.8 &128.7   &181.5 \\ \hline
\end{tabular}
\label{Table}
\end{table}

\begin{figure}
\centering
\subfloat[Kernel approximation.]{\includegraphics[width=0.45 \linewidth]{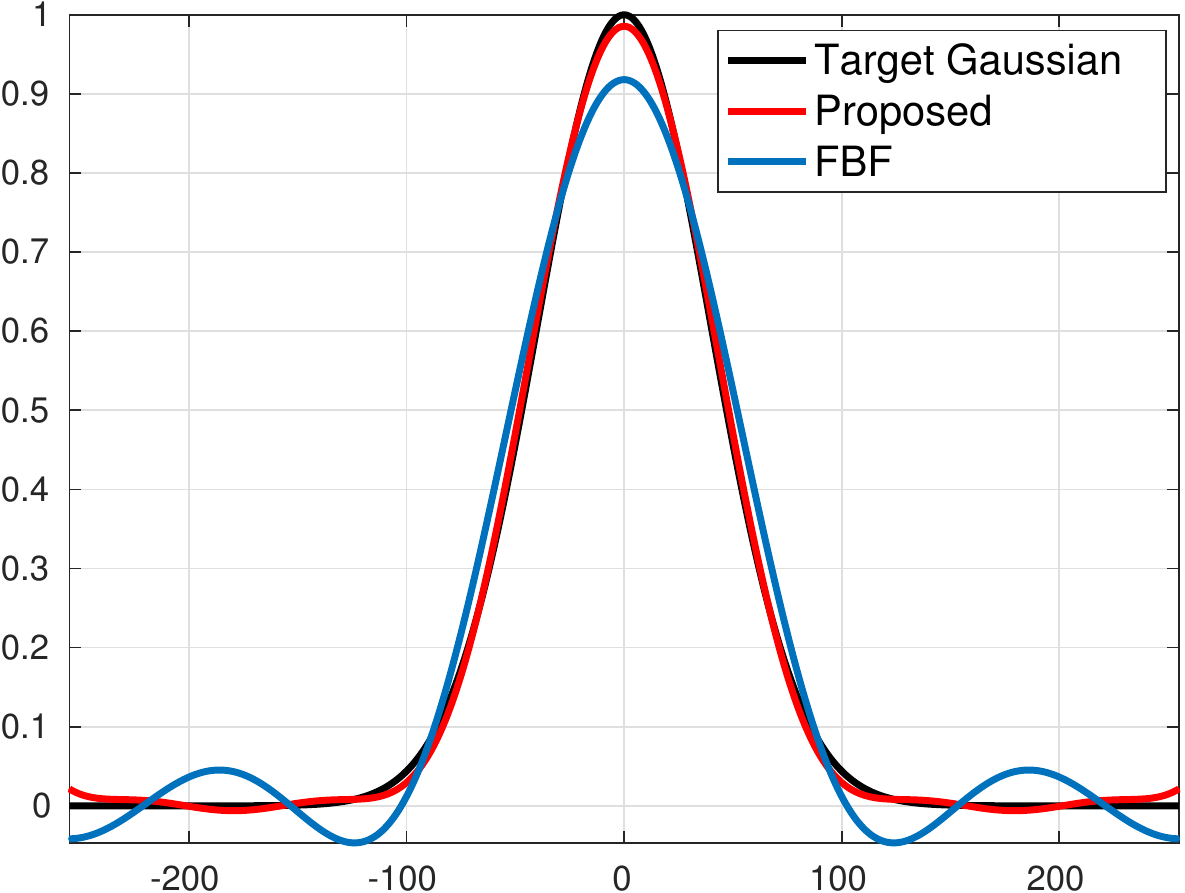}} \hfill
\subfloat[Error.]{\includegraphics[width= 0.45 \linewidth]{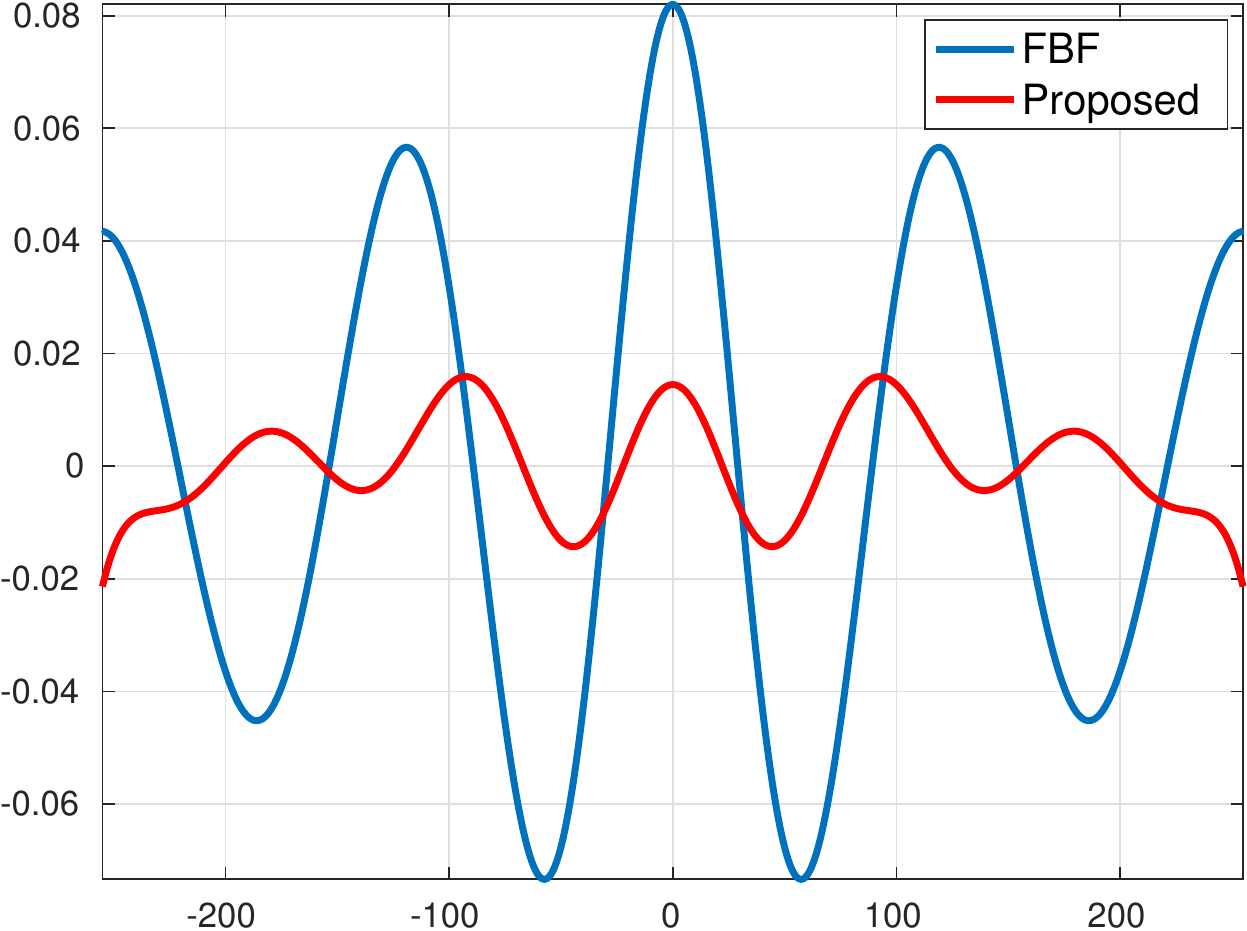}} 
\caption{Approximation of a Gaussian kernel ($\sigma=40$) using Fourier Bilateral Filtering (FBF) \cite{GhoshSPL2016} and our method. (a) Target kernel and its approximation; (b) Error between the target kernel and its approximation. The order is $K = 4$ in both  cases; for our method, we tuned $\varepsilon$ in Algorithm \ref{algo} to fix $K$.}
\label{fig:kernel_and_error}
\end{figure}

\begin{figure}
\centering
{\includegraphics[width=0.5\linewidth]{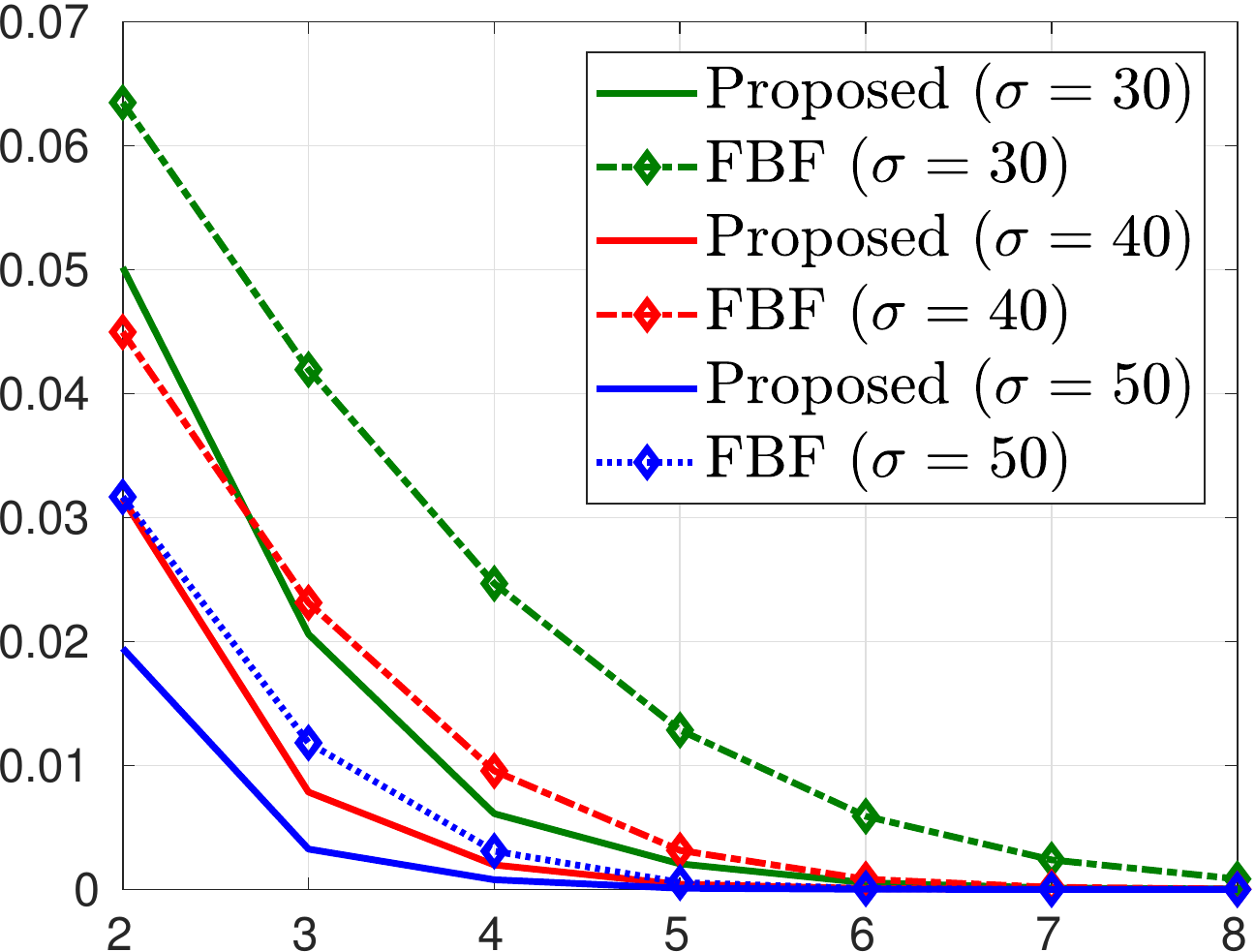}}
\caption{Comparison of the kernel error ($y$-axis) of FBF \cite{GhoshSPL2016} and the proposed method at different orders ($x$-axis). The kernel error in question is the mean-squared error between the samples of the target kernel and its approximation.}
\label{fig:ErrDecay}
\end{figure}

We first compare the proposed approximation with Fourier Bilateral Filtering (FBF) \cite{GhoshSPL2016}. 
The order for the former was fixed by adjusting $\varepsilon$ in Algorithm \ref{algo}; the same order was then used in FBF. 
A particular result is shown in Figure \ref{fig:kernel_and_error}, where we compare different approximations (of identical order) with the target kernel.
Notice that our approximation is much better than that of FBF. In particular, by optimizing the period, we are able to suppress the oscillation on the tail appearing in the FBF kernel. The approximation around the origin is also better for our method. In  { Figure \ref{fig:ErrDecay},} we show the decay of kernel error with the approximation order for both methods. As expected, the error is consistently lower with our method for different values of $\sigma$.

\begin{figure*}[!htp]
\centering
\subfloat[Input ($256 \times 256$).]{\includegraphics[width=0.235\linewidth]{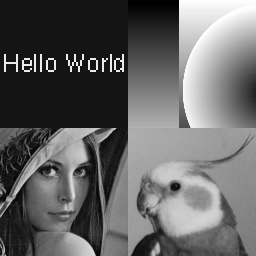}} \hspace{0.1mm}
\subfloat[FBF \cite{GhoshSPL2016}, PSNR = $68$ dB.]{\includegraphics[width=0.235\linewidth]{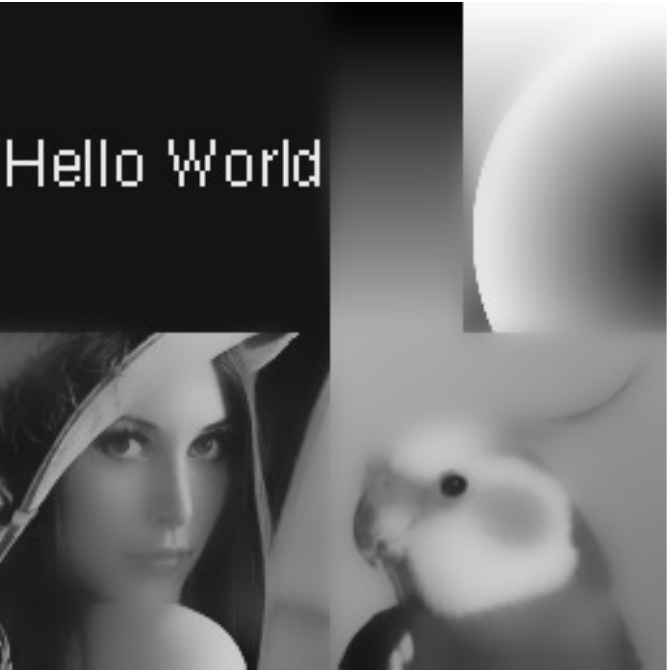}}\hspace{0.1mm}
\subfloat[CBF \cite{Kamata2015}, PSNR= $78$ dB.]{\includegraphics[width=0.235\linewidth]{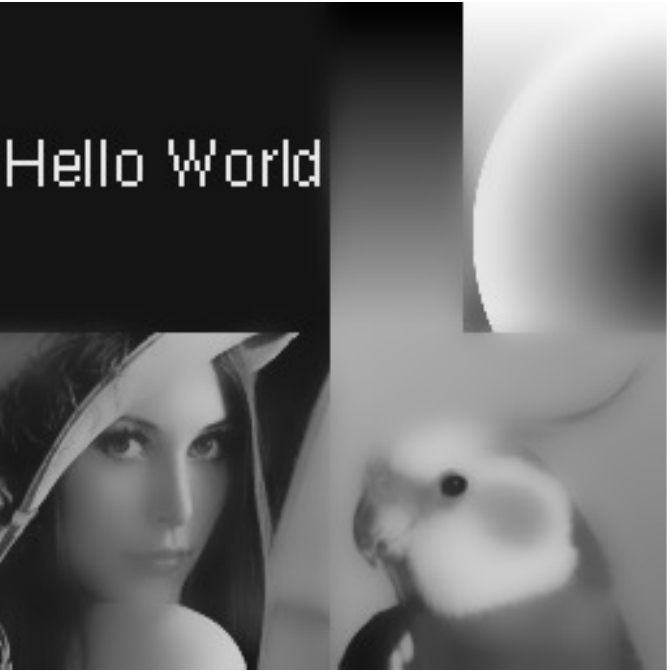}}\hspace{0.1mm}
\subfloat[Proposed, PSNR = $88$ dB.]{\includegraphics[width=0.235\linewidth]{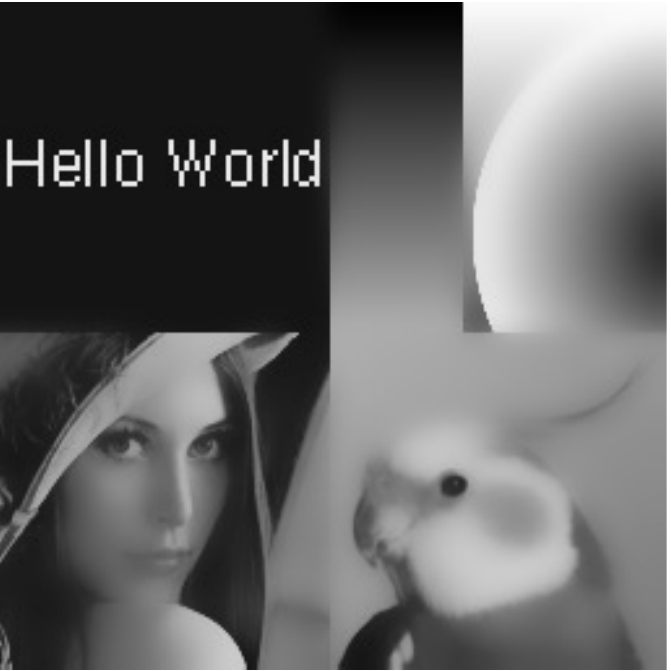}} \hspace{0.1mm}

\subfloat[Brute-force filtering.]{\includegraphics[width=0.235\linewidth]{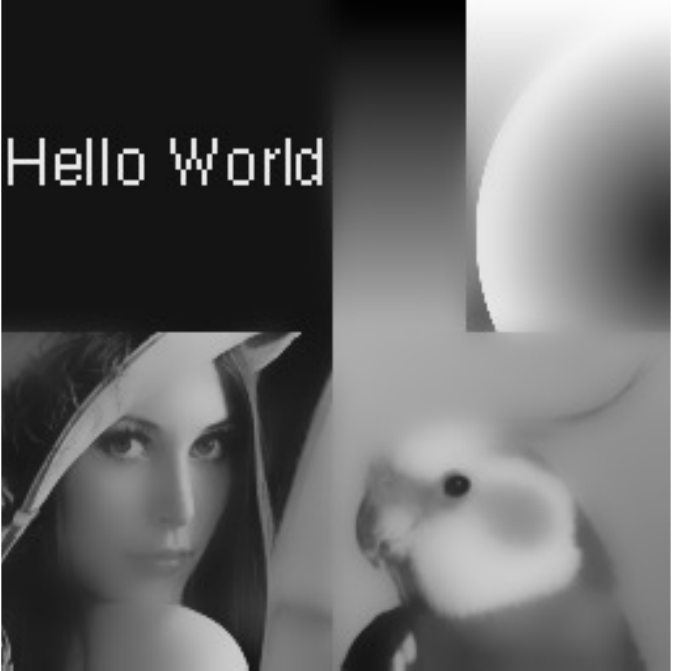}} \hspace{0.2mm}
\subfloat[Error  (e) - (b).]{\includegraphics[width=0.235\linewidth,height=0.23\linewidth]{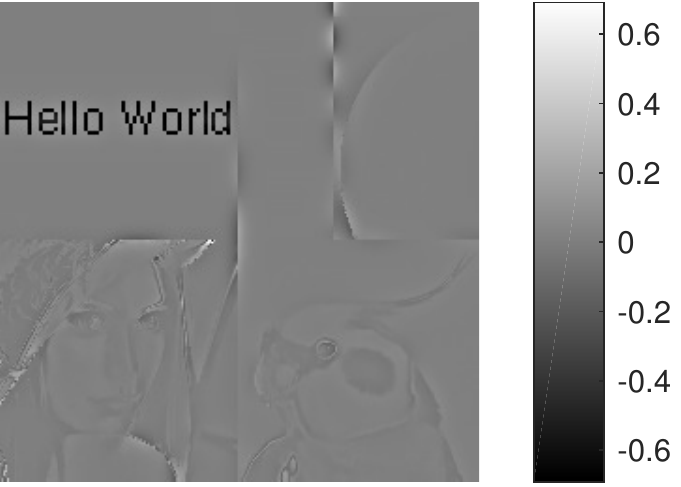}}\hspace{0.1mm}
\subfloat[Error (e) - (c).]{\includegraphics[width=0.235\linewidth,height=0.23\linewidth]{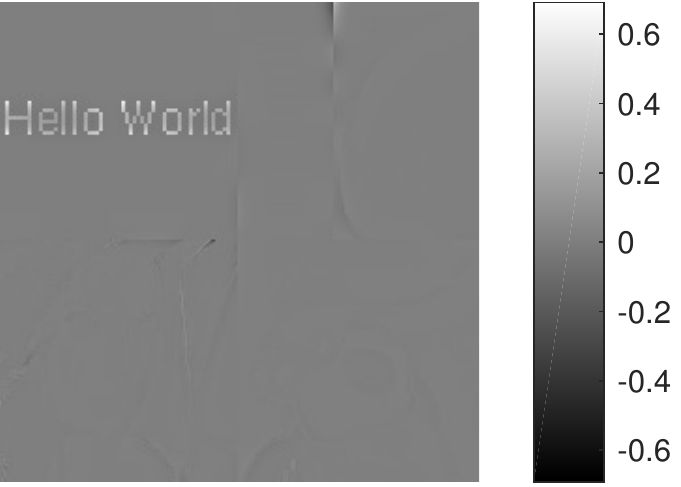}}\hspace{0.1mm}
\subfloat[Error (e) - (d).]{\includegraphics[width=0.235\linewidth,height=0.23\linewidth]{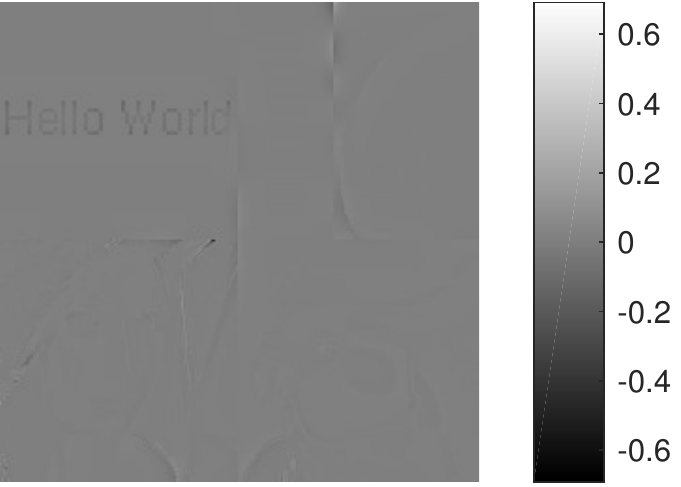}} \hspace{0.1mm}
\\~\\
\caption{Visual comparison of FBF \cite{GhoshSPL2016}, CBF \cite{Kamata2015}, and the proposed method for fixed order $K=5$. The error between the brute-force implementation of \eqref{BF} and the respective approximations are also shown. The parameters used are $\sigma = 55$ and $\theta = 5$. The corresponding PSNR value is mentioned in the caption.}
\label{Visualfig1}
\end{figure*}

\begin{figure*}[!htp]
\centering
\subfloat[Input ($256 \times 256$).]{\includegraphics[width=0.197\linewidth]{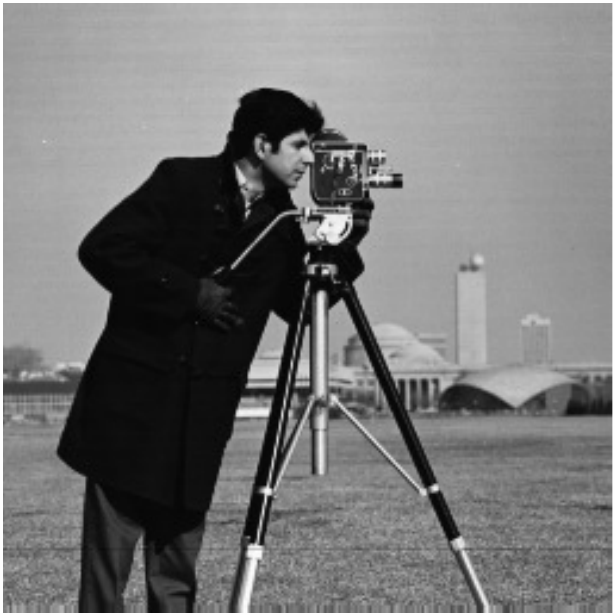}} \hfill
\subfloat[Brute-force filtering.]{\includegraphics[width=0.197\linewidth]{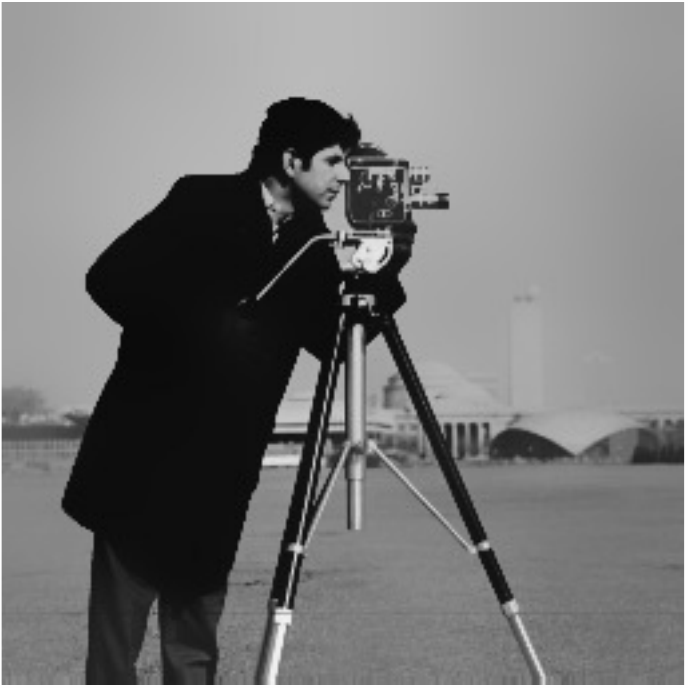}}\hfill
\subfloat[FBF \cite{GhoshSPL2016}, PSNR = $36$ dB.]{\includegraphics[width=0.197\linewidth]{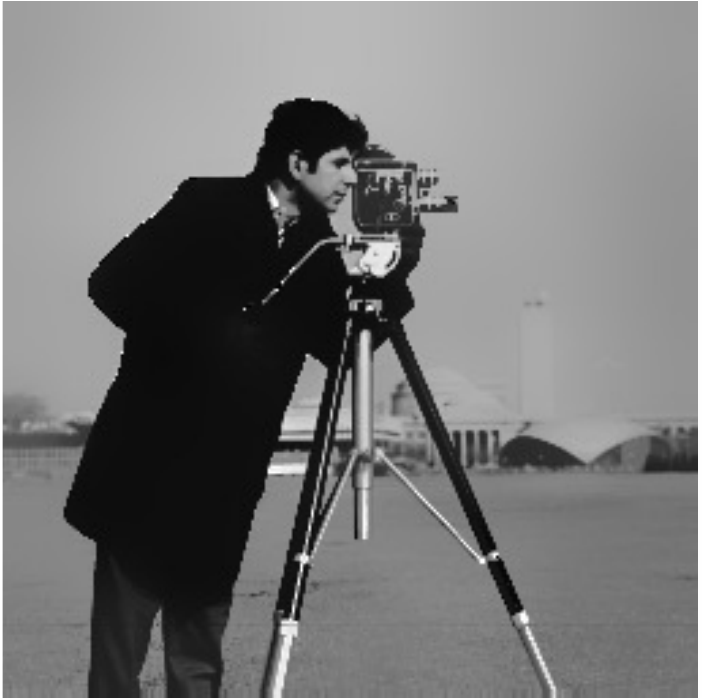}} \hfill
\subfloat[CBF \cite{Kamata2015}, PSNR = $52$ dB.]{\includegraphics[width=0.197\linewidth]{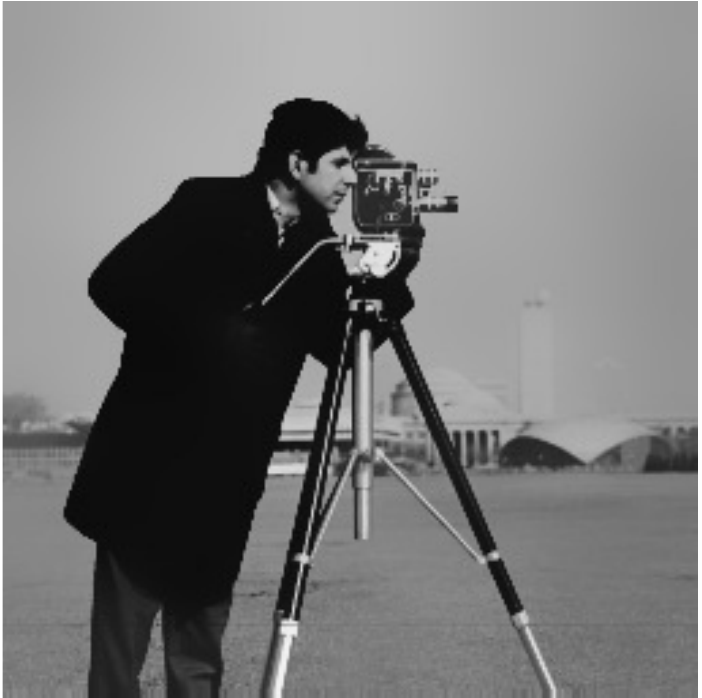}} \hfill
\subfloat[Proposed, PSNR = $55$ dB.]{\includegraphics[width=0.197\linewidth]{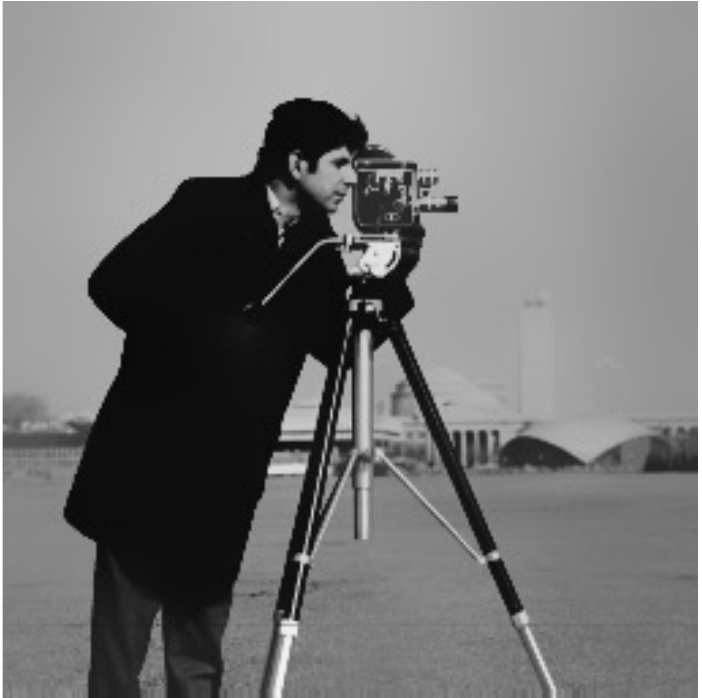}}
\\~\\
\caption{Visual comparison of FBF \cite{GhoshSPL2016}, CBF \cite{Kamata2015} and the proposed method for $K=5$. The parameters used are $\sigma = 30$ and $\theta = 10$.}
\label{Visualfig3}
\end{figure*}

Following existing works \cite{Durand2002,Yang2009,Mozerov2015}, we measure the filtering accuracy using the peak-signal-to-noise ratio: 
$\mathrm{PSNR} = 10 \log_{10}( 255^2/\mathrm{MSE})$,
where $\mathrm{MSE}$ is the mean-squared error between the brute-force and the fast approximation of \eqref{BF}. 
The timing and PSNR of the proposed approximation for different values of $\sigma$ and $\varepsilon$ are shown in Table \ref{Table}.
Notice that the timing scales linearly with $\log(1/\varepsilon)$.
For fixed $\varepsilon$, the timing (order) is more when $\sigma$ is small. This is generally the case with Fourier approximations \cite{Chaudhury2011a,Kamata2015,GhoshSPL2016}---it it is difficult to approximate a narrow Gaussian pulse using sinusoids.
On the other hand, our PSNR is consistently larger than the acceptable threshold of $40$ dB \cite{Porikli2008,Mozerov2015}, even when $\varepsilon$ is as large as $0.1$.

\begin{table}
\setlength{\tabcolsep}{4.0pt}
\centering
\caption{Average PSNR for images from the BM3D dataset \cite{ImgDatabase_bm3d}}
\label{table}
\begin{tabular}{|c|| c c c ||c c c|}
 \hline
Method \textbackslash{} $(\theta, \sigma)$   & $(2, 20)$  & $(2, 30)$  & $(2, 50)$  &  $(5, 20)$  &  $5, 30)$ &  $(5, 50)$ \\ \hline \hline
FBF \cite{GhoshSPL2016}  &37.3  &43.0  &46.5  &33.3  &38.7  &42.5 \\ \hline
CBF \cite{Kamata2015}    &44.9  &56.0  &58.5  &39.5  &51.1  &54.8 \\ \hline
Proposed    		 &45.6  &56.7  &59.8  &40.7  &52.0  &55.6 \\ \hline
\end{tabular}
\label{Table_avgPSNR}
\end{table}
We next compare the filtering performance with CBF \cite{Kamata2015}.
For a fair comparison, we have used the same order (number of convolutions) for all three methods.
That is, we kept the timings same and compared the PSNRs.
First, we set the order for our method using $\varepsilon$ in Algorithm \ref{algo}.
We next tuned the tolerance parameter $\tau$ in CBF to obtain the same order.
The order in FBF was set directly, since $T$ is not optimized in this case.
A couple of comparisons are shown in  {Figures \ref{Visualfig1} and \ref{Visualfig3}} on the grayscale images \textit{Montage} and \textit{Cameraman} \cite{ImgDatabase_bm3d}. 
For the example in Figure \ref{Visualfig1}, the PSNR improvement over CBF is about $10$ dB. 
While it is somewhat difficult to access this improvement by directly comparing the filtered images, it is evident from the  respective error images that the approximation is better near edges in our method. 
 {Considering the state-of-the-art performance of CBF, the $3$ dB improvement in Figure \ref{Visualfig3} is significant.
An additional visual comparison at $\sigma=25$ is provided in the \textit{supplement}, where the increment is by $4.2$ dB.
Finally, in Table \ref{Table_avgPSNR}, we have compared the average PSNRs over a set of images \cite{ImgDatabase_bm3d} for practical settings of $\theta$ and $\sigma$. Notice that our method is at least as good as CBF in terms of filtering accuracy. The PSNR improvement is in the range $0.7$ to $1.3$ dB. }

\section{Conclusion}
\label{Conc}

 {We showed that by jointly optimizing the order, period, and coefficients in \cite{GhoshSPL2016}, we can outperform the state-of-the-art CBF in certain cases. In general, we showed that the accuracy of our method is at least as good as CBF.
As mentioned in the introduction, unlike CBF, the proposed approximation can also be used for non-Gaussian range kernels \cite{Gunturk2011,Mirbach2012,Yang2014}. Also, we were able to establish convergence and provide a bound on the filtering error.}

\bibliographystyle{IEEEtran}

\begin{thebibliography}{9}


\bibitem{Tomasi1998} C. Tomasi and R. Manduchi, ``Bilateral filtering for gray and color images,'' \textit{Proc. IEEE International Conference on Computer Vision}, pp. 839-846, 1998.
 
\bibitem{Paris2009} S. Paris, P. Kornprobst, J. Tumblin, and F. Durand, \textit{Bilateral Filtering: Theory and Applications}, Now Publishers Inc., 2009.

\bibitem{Durand2002} F. Durand and J. Dorsey. ``Fast bilateral filtering for the display of high-dynamic-range images,'' \textit{ACM Transactions on Graphics}, vol. 21, no. 3, pp. 257-266, 2002.

\bibitem{Paris2006} S. Paris and F. Durand, ``A fast approximation of the bilateral filter using a signal processing approach,'' \textit{Proc. European Conference on Computer Vision}, pp. 568-580, 2006.

\bibitem{Weiss2006} B. Weiss, ``Fast median and bilateral filtering,'' \textit{Proc. ACM Siggraph}, vol. 25, pp. 519-526, 2006.

\bibitem{Porikli2008} F. Porikli, ``Constant time $O (1)$ bilateral filtering,'' \textit{Proc. IEEE Conference on Computer Vision and Pattern Recognition}, pp. 1-8, 2008.

\bibitem{Yang2009} Q. Yang, K. H. Tan, and N. Ahuja, ``Real-time $O (1)$ bilateral filtering,'' \textit{Proc. IEEE Conference on  Computer Vision and Pattern Recognition}, pp. 557-564, 2009. 

\bibitem{Chaudhury2011a} K. N. Chaudhury, D. Sage, and M. Unser, ``Fast $O(1)$ bilateral filtering using trigonometric range kernels,'' \textit{IEEE Transactions on Image Processing}, vol. 20, no. 12, pp. 3376-3382, 2011.


\bibitem{Gunturk2011} B. K. Gunturk, ``Fast bilateral filter with arbitrary range and domain kernels," \textit{IEEE Transactions on Image Processing}, vol. 20, no. 9, pp. 2690-2696, 2011.

\bibitem{Kamata2015} K. Sugimoto and S. I. Kamata, ``Compressive bilateral filtering," \textit{IEEE Transactions on Image Processing}, vol. 24, no. 11, pp. 3357-3369, 2015.

\bibitem{Mirbach2012} K. Al-Ismaeil, D. Aouada, B. Ottersten, and B. Mirbach, ``Bilateral filter evaluation based on exponential kernels," \textit{Proc. International Conference on Pattern Recognition}, pp. 258-261, 2012.
 
\bibitem{Yang2014} Q. Yang, ``Hardware-efficient bilateral filtering for stereo matching," \textit{IEEE Transactions on Pattern Analysis and Machine Intelligence}, vol. 36, no. 5, pp.1026-1032, 2014.

\bibitem{Mozerov2015} M. G. Mozerov and J. van de Weijer, ``Global color sparseness and a local statistics prior for fast bilateral filtering,'' \textit{IEEE Transactions on Image Processing}, vol. 24, no. 12, pp. 5842-5853, 2015.

\bibitem{GhoshSPL2016} S. Ghosh and K. N. Chaudhury, ``On fast bilateral filtering using Fourier kernels,'' \textit{IEEE Signal Processing Letters,} vol. 23, no. 5, pp. 570-573, 2016.

\bibitem{Dai2016}  L. Dai, M. Yuan, and X. Zhang, ``Speeding up the bilateral filter: A joint acceleration way,'' \textit{IEEE Transactions on Image Processing}, vol. 25, no. 6, pp. 2657-2672, 2016.
 {
\bibitem{Sugimoto2016} K. Sugimoto,  T. Breckon, and S. Kamata, ``Constant-time bilateral filter using spectral decomposition,'' \textit{Proc. IEEE International Conference on Image Processing}, pp. 3319-3323, 2016.}

\bibitem{Papari2017}  G. Papari, N. Idowu, and T. Varslot, ``Fast bilateral filtering for denoising large 3D images,'' \textit{IEEE Transactions on Image Processing}, vol. 26, no. 1, pp. 251-261, 2017.

\bibitem{Nair2017} P. Nair, A. Popli, and K. N. Chaudhury, ``A fast approximation of the bilateral filter using the discrete Fourier transform,'' \textit{Image Processing On Line,} vol. 7, pp. 115-130, 2017.

\bibitem{Gabiger2014}  A. Gabiger-Rose, M. Kube, R. Weigel, and R. Rose, ``An FPGA-based fully synchronized design of a bilateral filter for real-time image denoising,'' \textit{IEEE Transactions on Industrial Electronics}, vol. 61, no. 8, pp. 4093-4104, 2014.

 {\bibitem{ImgDatabase_bm3d} BM3D Image Database, \url{http://www.cs.tut.fi/~foi/GCF-BM3D/BM3D_images.zip}. }


\end{thebibliography}

\end{document}